\documentstyle[aaspptwo]{article}

\journalid{337}{1 September 1993}
\articleid{11}{14}

\begin{document}

\title{Threshold Star Formation Effects in the Peculiar Galaxy \\
Arp~10 (=~VV 362)}

\author{V. Charmandaris\altaffilmark{1} and P. N. Appleton\altaffilmark{2}}
\authoraddr{Dept. of Physics and Astronomy, Iowa State University, Ames,
           IA 50011}
\affil{Erwin W. Fick Observatory \\
Department of Physics and Astronomy, Iowa State University, Ames, IA 50011}

\and

\author{A. P. Marston\altaffilmark{2}}
\affil{Department of Physics and Astronomy, Drake University,
Des~Moines, IA 50311 }

\altaffiltext{1}{Send offprint requests to V. Charmandaris.
                 e-mail: {\em vassilis@iastate.edu}}

\altaffiltext{2}{Visiting astronomer at the Kitt Peak National Observatory.
KPNO is operated by the Association of Universities for Research in
Astronomy, Inc., under contract with the National Science Foundation.}

\vspace{1cm}
\begin{center}
To appear in The  Astrophysical Journal, September 1, 1993.
\end{center}

\begin{abstract}
We present images of the peculiar galaxy Arp~10 which reveal two rings
of concurrent star formation.  Apart from a bright ring visible on
early photographs of the system, an even brighter inner ring of
\ion{H}{2} regions is found within the nuclear bulge of the galaxy.  A
very faint ring-arc of \ion{H}{2} regions is also seen associated with
a third outer ring or shell.  We detect a small companion galaxy on
the minor axis of Arp~10 and an extended extra-nuclear knot which lies
between the first and second rings.  An investigation of the H$\alpha$
fluxes in the rings reveals an increase in the emission where the
ring surface density in the R-band light exceeds
21.3$\pm$0.2 mag\,arcsec$^{-2}$.  If the R-band light is dominated by
old stars in the underlying density wave, then the results suggest
evidence for a star formation law which exhibits a threshold dependence
on the strength of the density wave in the rings. Even if the R-band
continuum in the ring is heavily contaminated with red light from the
underlying young stars (a result at odds with the smooth continuum
morphology of the ring) then a smaller, but still significant
non-linear enhancement in the star formation rates in one segment of
the second ring is required to explain the results.  In either case,
strong trends in the star formation rate with azimuth around Ring~2
are in good agreement with off-center collisional ring galaxy models.
\end{abstract}


\clearpage

\section{Introduction}

Arp~10 (=~VV\,362) is a galaxy containing a bright ring, an off-center
nucleus and a faint bar. Vorontsov-Velyaminov (1977) describes the
galaxy as containing `` \ldots an enormously developed massive part of
the ring'' which we will later show is a region of enhanced star
formation in the ring.  Arp 10 was noted as a ring galaxy by Dahari
(1985).  Faint traces of outer filaments resembling antennae are seen
protruding from the galaxy in the photograph presented in the Arp
Atlas of Peculiar Galaxies (Arp 1966) suggesting that Arp~10 may be a
colliding system. Although the outer filaments in Arp~10 in some ways
resemble shells around some elliptical galaxies (Malin and Carter
1983), the single dish \ion{H}{1} spectrum of the galaxy obtained by
Sulentic and Arp (1983) exhibits the typical two-horned profile of a
rotating planar disk.

We present deep CCD images of Arp~10 which support the collisional
picture for the formation of the ring and filaments. A small
elliptical companion is found approximately one ring diameter from Arp
10 on its minor axis, suggesting that Arp~10 may be a ring galaxy
formed by the head-on collision of a small intruder galaxy through the
center of a rotating disk (Lynds and Toomre 1976; Theys and Spiegel
1977). Another ring and additional ring-arcs are also discovered in
the light of H$\alpha$ in Arp~10. The newly discovered ring lies in
the inner regions of the central bulge. The H$\alpha$ imaging also
reveals sharply defined ring-arcs associated with the faint outer
filaments. Multiple rings are expected in well evolved models of
colliding ring galaxies (Toomre 1978; Appleton and Struck-Marcell
1987, hereafter ASM; Struck-Marcell 1990).

Ring galaxies provide an ideal laboratory for exploring models of star
formation triggered by density waves.  Of special interest is the
mildly off-center collision between an intruder galaxy and a disk. It
has been argued by Appleton and Struck-Marcell (1987) that such a
collision will generate an expanding ring wave which shows large but
smooth variation in stellar and gaseous density around the ring. Such
a wave is an ideal tool for exploring the dependence of star formation
rate (hereafter SFR) on ring surface density in galaxies. The work is
relevant to threshold star formation mechanisms such as those put
forward by Scalo and Struck-Marcell (1984; 1986), Struck-Marcell and
Appleton (1987), hereafter SMA, for colliding systems as well as
Kennicutt (1989; 1990) for normal galaxies (See also observations by
Skillman 1987; Schombert and Bothun 1988; Carignan and Beaulieu 1989).

In this paper we will investigate the peculiar morphology of Arp~10
and will calculate the SFR in the bright knots in all three ring
structures. In \S~2 we describe our observations and we present the
optical morphology of Arp~10 in \S~3. In \S~4 we present details of
how we performed the photometry and the star formation rate (SFR)
calculations. The results are presented in \S~5.  In \S~6 we
describe some predictions based on existing models.  In \S~7 we
discuss their physical significance, and our conclusions are
highlighted in \S~8.  The systemic heliocentric velocity of Arp~10 is
9093 km\,s$^{-1}$ (Sulentic and Arp 1983). Using the velocity
corrected for motion relative to the local group (de Vaucouleurs de
Vaucouleurs and Corwin 1976) and assuming a value for the Hubble
constant of 75 km\,s$^{-1}$\,Mpc$^{-1}$ we estimate a distance to
Arp~10 of 122 Mpc.

\section{Observations}

The observations were made during photometric conditions on the night
of January 19 1991 using the KPNO 2.1~m telescope. Images were
obtained through a broad-band R and a narrow band 80 \AA\ wide filter,
centered close to the wavelength of redshifted H$\alpha$ and
[\ion{N}{2}].  The detector was a Tek2 512$\times$512 pixel$^{2}$ CCD.
Two 1500~s observations were made using the H$\alpha$ + [\ion{N}{2}]
filter in order to detect emission from the star forming knots.  A
1200~s exposure was taken using the R-band filter. Bias subtraction
and flat fielding of the images was performed in a standard way. Flat
fields were obtained using sky observation made during the morning
twilight. The continuum was removed from the spectral line images
using a carefully scaled R-band image. The seeing disks were very
similar for both the line and continuum images (FWHM = 1.2 arcsec).
Calibration of the photometry was performed using data taken from the
star Hz 15 (Stone 1974) and the standard star 95 52 (Landolt 1983).

As a further check of the reliability of the R-band image as a good
representation of the stellar continuum, we made further observations
of Arp~10 with the newly commissioned wide-field CCD system at Iowa
State University's E. W. Fick Observatory.  The Fick wide-field CCD
camera was attached to the 0.6~m Mather telescope with a focal reducer
providing an f/4 beam at the detector. The CCD is a thinned TI
800$\times$800 pixel$^{2}$ device providing a 1.3 arcsec\,pixel$^{-1}$
image scale. The observations of Arp~10 were made during the nights of
August 28 and 29 1992 through a narrow band, 50 \AA\ wide interference
filter, centered at 6700~\AA.\ This provided a narrow-band continuum
with no possibility of contamination by hydrogen emission lines.  No
discernible difference was found between the KPNO R-band and the Fick
Observatory image down to a level of 5\%. (see Appendix)

\section{Optical Morphology}

A grey-scale representation of the R-band image is shown in Figure~1.
The galaxy has a dominant ring containing a nucleus and faint bar and
an outer structure which extends from the south-east in shell-like
filaments.  We indicate in Figure~1 a possible elliptical companion
galaxy to Arp~10, which lies 60 arcsec to the north-east on the minor
axis of the ring.  If Arp~10 is a collisional ring system this is a
prime candidate for the intruder galaxy. A bright extra-nuclear knot
is also seen close to the nucleus of Arp~10 to the south-west. This
knot, which is also marked in Figure~1, may be in some way related to
the collisional nature of the system. It is also visible in H$\alpha$
emission, which indicates that is must belong to the system. We will
not elaborate more on its nature since we do not yet have radial
velocity measurements which would help to further define its
relationship to the disk of Arp~10.

In Figures 2 and 3 we show a grey-scale image and a contour map, of
the H$\alpha$+[\ion{N}{2}] line emission from Arp~10. In addition to
the bright ring we also see for the first time the strong inner
line-emitting ring which lies within the central bulge of Arp~10.
Faint outer emission knots are also visible associated with the
shell-like continuum especially to the north-east.  We will refer to
the inner ring, the bright intermediate ring and the outer ring-arcs
(see Figure~2) as Rings~1, 2 and  Ring-arc~3 throughout this
paper.  The radii of the three ring-like structures are 2.8, 21.5 and
40 arcsec respectively.

\section{Photometry and SFR Calculations of the Ring Knots and Arcs}

The photometric fluxes of the many bright knots in Arp~10 were
determined in two ways:

\begin{itemize}
\item I.  We integrated the flux contained within each knot down to an
H$\alpha$ isophotal level of $0.60\times 10^{-16}$ ergs\,
cm$^{-2}$\,s$^{-1}$\,arcsec$^{-2}$\ (2.5 $\times$\ rms noise level in
the image).  The boundaries defined by this isophotal level in the
H$\alpha$ map were then transferred to the suitably registered R-band
continuum image in order that the R-band flux be estimated over the
same area.  From these flux estimates and a knowledge of the area of
the limiting contour, we then calculated the surface brightness of the
regions in both H$\alpha$ and red continuum light.

\item II. We performed photometry in a fixed circular software aperture
with a radius of 2.87 arcsec centered on each bright knot of Ring~2.
Again the surface brightness at the same aperture center was estimated
once the fluxes had been evaluated.
\end{itemize}

We calculated of the surface brightness using these two different
methods in order to ensure that the surface area of the flux
evaluation did not play a significant role in the final results.
Indeed, as will be shown below, the two methods yield similar results.

In Figure~2, we indicate the specific areas over which the flux was
evaluated down to the isophotal limit given above (Method~I). The
regions, presented in Table~1, are labeled 1--29, for future
reference. We note that for the very bright north-west region of
Ring~2, an entire portion of the ring exceeds the isophotal limit
because the many bright knots which lie close together in that region
cause the contours to merge. Since we are primarily interested in
surface brightness rather than total flux in our discussion below, we
decided to split up this part of the ring into segments equal in
length to the isophotal thickness of the ring (about 5 arcsec). We
attempted to include one bright \ion{H}{2} region complex in each ring
segment and avoided bisecting bright partially resolved hot-spots. A
similar argument applies to the inner ring which also exceeds the
isophotal limit over its entire extent. The two western knots
associated with the ring-arc were included in the calculations.
However, one section of the outer ring-arc subtends a large area at a
surface-brightness level of 0.3$\times 10^{-16}$ ergs\,s$^{-1}$\,
cm$^{-2}$. Although this was below our isophotal limit, we decided to
include the arc region for completeness.

The regions used in the constant aperture method (Method~II) a--s, are
presented in Table~2. The circular aperture was centered at each
bright knot which was included in the regions selected for Method~I.
In this way the regions a--f correspond roughly to regions 6--11 and
regions g--s to regions 13--25.

Star formation rates have been estimated from H$\alpha$ fluxes by a
number of authors but most notably by Kennicutt (1983) and Gallagher
Hunter and Tutukov 1984 (hereafter GHT).  The assumptions in either
case are similar. It is usually assumed that the H$\alpha$ emission
arises from Case~B hydrogen recombination in balance with a strong
ionising continuum dominated by stars in the 30 to 60 $M_{\sun}$
range.  Observationally, the H$\alpha$ flux can be converted to the
number of Lyman continuum photons through the expression :

\begin{equation}
N_{c} = 8.78\times10^{61}F_{c}(\mbox{H}\alpha) D^{2}
\end{equation}

where D is the distance to the galaxy in Mpc, and
$F_{c}(\mbox{H}\alpha)$ is the H$\alpha$ flux in units of
ergs\,s$^{-1}$\,cm$^{-2}$ corrected for reddening.  In order to derive
from this a SFR it is usual to assume a fixed slope for the IMF
and an upper mass cutoff. If these factors do not change from
region to region in a galaxy and if, additionally, there are few or no
super-massive stars present in the \ion{H}{2} regions (e.g. Wolf-Rayet
stars), then changes in H$\alpha$ flux are attributed to changes in
the total number of stars born per year.  For example, GHT estimate
that for a Salpeter IMF of the form $\Phi(M)=M^{-a}$ with $a=2.35$ and
$M_{upper}= 100\,M_{\sun}$, the SFR is :

\begin{equation}
\alpha_{c} = 2.5\times10^{-54} N_{c}\ \mbox{stars\,yr}^{-1}
\end{equation}

Virtually all the ionising photons are produced by 30--60 $M_{\sun}$
stars with lifetimes $3\times10^{6}$ yr; so formally $\alpha_{c}$
provides an instantaneous value of the SFR. Integrating over the mass
range 10--100 $M_{\sun}$, it is easy to show that this corresponds to
a SFR, measured in gas mass converted into massive (M $> 10 M_{\sun}$)
stars per year, of 0.7$\alpha_{c}$ $M_{\sun}$\,yr$^{-1}$.  If all
stars are included down to a lower mass cut-off of 0.1 $M_{\sun}$ the
total SFR would equal 5.8$\alpha_{c}$ $M_{\sun}$\,yr$^{-1}$. This
agrees closely with the estimate by Kennicutt (1983) adopting very
similar assumptions.


\section{Results}

The results of Method~I are presented in Table~1. Column (1) gives the
name of the region identified in Figure~2 and Column (2) presents the
area of the region in arcsec$^{2}$. In Columns (3) and (4) we show the
magnitude of the region measured through the H$\alpha$ and R-band
filters respectively (see \S~2). In Column (5) we give the H$\alpha$
surface brightness, after applying a correction for an assumed 5\%
contamination of the H$\alpha$ flux by [\ion{N}{2}] emission.  Such a
correction is consistent with the work of Kennicutt~ (1983) on
irregular galaxies and is similar to that measured in the Cartwheel
ring galaxy by Fosbury and Howarden (1976). In Column (6) we present
the R-band surface brightness of each region.  Both columns (5) and
(6) have also been corrected for a Galactic extinction of A$_{v}=0.15$
(de Vaucouleurs, de Vaucouleurs and Corwin 1976). In Column (7) we
show the luminosity of each region and finally in Column (8) we
present the SFR in units of stars\,Gyr$^{-1}$\,pc$^{-2}$ . We note
that no correction has been made for internal extinction within the
galaxy itself.

In Table~2 we present the results of Method~II. Column (1) indicates
the region identified in Figure~2, Column (2) presents the
constant area in arcsec$^{2}$, and the Columns (3) to (6) have similar
information to that of Table~1.

The luminosities of the \ion{H}{2} region complexes (see Table~1) show
a wide spread in values, ranging from approximately 3$\times$10$^{39}$
ergs\,sec$^{-1}$ in the outer ring-arc (region 27) to
2$\times$10$^{40}$ ergs\,sec$^{-1}$ in the inner ring (region 2).
Although the regions we have evaluated probably contain more than one
individual \ion{H}{2} region (this is especially true in the bright
north-western regions of Ring~2), it is interesting to note that in
all cases the luminosities are comparable with the brightest
\ion{H}{2} regions seen in late-type galaxies (Kennicutt (1988)). For
example, even the emission from knot 28 in the outer ring-arc has the
luminosity of an average \ion{H}{2} region observed in an Sbc or Sc
galaxy.


\section{Model Predictions for Ring Galaxies}

It was demonstrated by Toomre (1978) that mildly off-center collisions
between galaxies drive ring-like waves through their disks. This work
was extended by ASM and SMA to include the study of the star formation
response in ring galaxies. This showed that the overdensity in the
ring can be used as a probe of the sensitivity of the star formation
rate to the density in the ring wave. Indeed, mildly off-center
collisions are shown to lead to a strong azimuthal variation in ring
gas density. In Figure~4a we show the density distribution of an
expanding ring wave driven into a disk derived from Figure~7c of ASM.
We have modified the figure to include a notation which will allow us
to discuss further the azimuthal dependence of the star formation rate
on the ring density, using the letters A to F to indicate azimuth
angles around the ring at intervals of 60 degrees. The dependence of
the predicted star formation rate on the ring density (relative to the
unperturbed density) is shown for two forms of star formation rate
law. The first, Figure~4b, is derived directly from ASM Model E and as
discussed in that paper, follows approximately a Schmidt law in which
the SFR $\propto$ $\rho^{1.5}$.  The second law, hereafter called the
``threshold'' model, has a SFR which follows a Schmidt law below some
critical gas surface density, but then increases rapidly above that
density (see Figure~4c).  Using the lettering system defined in
Figure~4a to identify azimuthal position of the region around the
ring, it is noted that both star formation laws exhibits systematic
behavior with azimuth.  In the case of the Schmidt law model, the
behavior in the SFR/density domain is a loop which begins at the left
of the diagram (Position A in Figure~4a; low density, small star
formation rate), rises monotonically through the densest part of the
wave (B through D) and then falls back to its initial value in the
least dense part of the ring (E through F). In the case of the
threshold model, similar behavior is found until the threshold is
exceed, where the SFR follows a much steeper path, again forming a
tight loop. The threshold model of Figure~4c, shows some important
differences from the Schmidt-type model of Figure~4b.  Firstly, the
SFR/density relationship shows a sudden discontinuity at the critical
density.  This would have direct observational consequences if star
formation rates and ring densities could be determined over a wide
enough dynamic range on either side of the discontinuity. Secondly,
because of the peculiar geometry of the off-center collision shown
here, the sudden steepening of the SFR law at the critical density
translates into a sudden brightening of the ring in H$\alpha$ light in
an arc encompassing the densest parts of the ring. For example, in
Figure~4c, the azimuthal regions of the ring extending between points
C and E in Figure~4a, lie well above threshold and would be emitting
strongly at the wavelength of H$\alpha$ (We note that point D is so
luminous in this model it lies outside the range of the plotted points
in Figure~4c). In contrast, from Figure~4b, it can be seen that only a
small region near point D would be expected to be producing stars at a
high rate. For the threshold models it is clear that the physical
extent of the bright region will be strongly dependent on the value of
the threshold and the strength of the density wave and will differ
from one galaxy to another. It is important to note that these
predictions relate only to gaseous density waves although it is
expected that at least initially, the amplitude of the stellar and
gaseous wave will be similar in the collisional picture. However, an
early-type disk galaxy containing no cool gas is unlikely to exhibit
the star formation behavior described here.

The threshold models predict that for an off-center collision, a high
contrast, high star formation rate segment of the ring will develop.
{}From Figure 2, it is likely that Arp~10's intermediate ring (Ring~2)
fits this description well. The north-western quadrant of the ring is
extremely luminous in H$\alpha$ light and this will be quantified in
the next section. We note also that the size of the \ion{H}{2} regions
grow suddenly in this region and this will be considered as further
evidence for threshold behavior.


\section{Evidence for a Star Formation Threshold in Arp~10 }

\subsection{The Surface Brightness Properties of the Rings}

In Figure~5 we present the H$\alpha$ surface density as a function of
the R-band continuum around the ring features. These data have been
obtained from values taken from Table~1. The knots associated with the
three ring structures are indicated. As one proceeds from Ring-arc~3
(outer ring-arc) to the slightly higher R-band surface brightness
levels, $\sigma_{R}$, in Ring~2, the H$\alpha$ surface brightness
increases slowly with an average value of approximately
3.1$\times$10$^{-22}$ ergs\,s$^{-1}$\,cm$^{-2}$\,pc$^{-2}$. However,
as one proceeds to higher levels of $\sigma_{R}$, an increase in
scatter of the points is observed near $\sigma_{R}$=21.3$\pm$0.2 mag\,
arcsec$^{-2}$. Although some of the scatter is attributable to
measurement error, (especially region 14), most of the scatter is due
to regions 19 through 23, which are significantly more luminous (by a
factor of 5 on a linear scale) in H$\alpha$ light than the other
regions with similar $\sigma_{R}$.  These points correspond to the
bright regions of the ring noted by Vorontsov-Velyaminov.  Finally, we
proceed to Ring~1 (the inner ring).  Here the H$\alpha$ surface
brightness is well determined but the strength of the red continuum is
only poorly known. This is because the ring is embedded in a strong
central bulge which dominates the red light. We therefore accounted
for the bulge contribution to the light by fitting an
R$^{\onequarter}$ law profile to the bulge and then removing its
contribution to the inner ring. Because of uncertainties in the fit to
the bulge light, the surface brightness at R-band for the Ring~1
points have correspondingly large errors (see Figure~5).  However,
they do provide an important extra piece of evidence for a threshold
star formation law in Arp~10 as discussed below and so we have
included them for completeness in Figure~5.

As discussed in \S~4, we repeated the surface brightness estimates
for the Ring 2 using the constant aperture method (Method~II). We
confirmed that an increase in dispersion was observed in the H$\alpha$
surface brightness at a level of 21.3 mag\,arcsec$^{-2}$,
in accord with the earlier result of Figure~5.

\subsection{The Possible Contamination of the R-band Light by Young Stars}

Before proceeding to discuss the implications of Figure~5 for testing
models of density wave induced star formation, we must first question
the assumption that the red continuum light is dominated by old stars
in the ring waves, rather than light from young star clusters. We have
already noted that the distribution of R-band emission in the rings is
rather smooth relative to that of the H$\alpha$ emission and this fact
alone argues that our continuum measurements are not strongly
influenced by light from young stars. However, for the purposes of the
following discussion we will adopt a ``worst-case contamination''
approach to the problem and assume that all of the red continuum light
results from emission from underlying young stars and explore its
impact on our conclusions.

If the underlying young clusters dominate the red continuum then this
is likely to lead to a correlation between log(F(H$\alpha$)) and a
broad-band continuum magnitude. Such a loose correlation is observed
in studies of \ion{H}{2} regions in late type galaxies (see Kennicutt
and Chu (1988)). The correlation can be understood in terms of a set
of self-similar clusters of increasing size in which the ratio of the
H$\alpha$ to continuum flux remains fixed. Assuming that all other
properties of the cluster remain the same (age, metallicity, IMF) then
such a set of clusters would lead to a linear correlation between
log(F(H$\alpha$)) and $\sigma_{R}$ with a slope of -0.4.  Indeed we
note that formally fitting a straight line to the data for Ring~2 and
Ring-arc~3 yields a line with a slope close to this value (dotted line
in Figure~5). In order to explore this possibility further, we show in
Figure~6, the residuals of Figure~5 after subtracting from it the
linear fit to the data.  To a first approximation, the scatter in the
residuals shown in Figure~6 might be taken to imply a good fit to the
data. However, a number of important points need to be emphasized.
The first is that the points 19 to 23 (corresponding to the bright
\ion{H}{2} regions in the north-western quadrant of Ring~2) lie
significantly above the linear relationship, given their small
observational errors. As previously stated, these points are a factor
of 5 times brighter than the average for their values of $\sigma_{R}$.
Secondly, we will show conclusively in \S~7.4, that the apparent
scatter in the residuals shown in Figure~6 for Ring~2, especially near
$\sigma_{R}$= 21.3$\pm$0.2 mag\,arcsec$^{-2}$, arises from {\em
systematic azimuthal changes} around the ring, not random errors in
the linear relationship.  This subtle point is crucial to our later
argument about the nature of the enhancement of emission in the
north-western quadrant of Ring~2. We will demonstrate below, that
these systematic variations away from the linear fit are naturally
explained within the context of the threshold star formation model.
However, as the linear fit to Figure~5 may imply, we cannot rule out
some contamination by young stars as an explanation for the increase
in red light in the brighter H$\alpha$ emitting regions. Even if some
degree of contamination does exist from young stars, it does not
invalidate the main conclusion, that the \ion{H}{2} regions observed
in regions 19--23 of Figure~2 are unusually bright as compared with
other \ion{H}{2} regions with the same R-band continuum.
\footnote {Another approach to the question of contamination of
the R-band flux from the underlying host cluster would have been to
predict the expected contribution to the R-band emission based on the
strength of the H$\alpha$ flux. In order to do this we would have had
to assume values for the IMF, the metallicity and the age of the
clusters and use a color evolution model to predict the expected
R-band continuum.  We believe that although such models exist for
solar abundances, the reliability of such models, especially the
treatment of the giant and supergiant stars is still quite
questionable. We therefore preferred to make use of the rough
correlation derived from observations.}

\subsection{Density Wave Induced Star Formation}

Another equally valid way of interpreting Figure~5 is to assume that
Arp~10 is a collisional ring galaxy in which density waves are driven
through its disk. The slow decline in $\sigma_{R}$ as one proceeds
from inner Ring~1 radially outward to the outer rings would be
explained as a consequence of the decreasing strength of the density
waves with radius. If we further assume that the stellar ring density
can be used to measure the gas density in the expanding wave (see
\S~7.6), then Figure~5 can be used as a diagnostic of the various forms
of star formation law discussed in \S~6. For example, deviations from
a simple linear relation in Figure~5, can be interpreted as deviations
from a Schmidt law behavior. We have already pointed out in \S~7.2,
that systematic deviations of this kind do exist.  We will now
investigate the possibility that these deviations are evidence for a
threshold star formation picture.

The evidence for a SFR threshold law in Arp~10 depends on two things.
The first is our ability to recognize a discontinuity in the SFR
versus ring density relationship consistent with our model predictions
and the second is that such a discontinuity should exhibit the
azimuthal variations predicted for the off-center collisional model.
As was stated earlier, the former problem is one of dynamic range.
Within any one ring, the dynamic range in the star formation rate is
relatively small, despite the apparently bright regions in Ring~2
discussed earlier. Arp 10 provides a useful probe of star formation
mechanisms because its three rings sample a wide range of densities,
from the weak outer Ring-arc~3 to the more overdense inner rings.

One simple method of finding a sudden discontinuity in the data
similar to that of Figure~4c is to split the data presented in
Figure~5 into two equal sets ordered by the H$\alpha$ flux and
independently fit a straight line to them. As long as the
discontinuity lies somewhere near the center of gravity of the data
then two lines of markedly different slope would be found. On the
other hand, if the same slope is found for both lines, then this would
argue for a continuous function of H$\alpha$ emission with continuum
light.  In Figure~5 we show the result of fitting two straight lines
to the data for all three rings using a Gauss-Markov elimination
method.  The method depends critically on good sampling of points on
either side of the discontinuity. The only restriction to the fitting
procedure was that the crossing point of the two lines should occur
somewhere within the body of these data. Despite the very uncertain
values for $\sigma_{R}$ in Ring~1 (giving these points lower weight),
there is evidence for a discontinuous form to the data presented in
Figure~5 since the fitted lines do have different slopes. The
intersection between the two lines occurs at around $\sigma_{R}$= 21.2
mag\,arcsec$^{-2}$ which is similar to the surface density at which
the increase in dispersion was noted in the Ring~2 emission alone. As
long as it is valid to treat all three rings together, the evidence
suggests a threshold star formation law similar to that discussed in
\S~6.

The increase in H$\alpha$ surface brightness at the R-band threshold
of 21.3$\pm$0.2 mag\,arcsec$^{-2}$ is most easily interpreted as a
threshold in the massive SFR. Using the values for the knots presented
in Table~1, we plot, in Figure~7, the SFR per square parsec as a
function of the red continuum surface density. The discontinuity in
H$\alpha$ surface brightness translates, under these assumptions, into
a sudden change in SFR.  The rates change from an average of 0.5--1.5
stars\,Gyr$^{-1}$\,pc$^{-2}$ (or 2.9--8.7
$M_{\sun}$\,Gyr$^{-1}$\,pc$^{-2}$) in the fainter outer ring-arc and
the faint southern sections of Ring~2 to values which rise to 2--6
stars\,Gyr$^{-1}$\,pc$^{-2}$\ (or 11--30
$M_{\sun}$\,Gyr$^{-1}$\,pc$^{-2}$) in the bright north-west portion of
Ring~2.  Higher star formation rates of approximately 20
stars\,Gyr$^{-1}$\,pc$^{-2}$ (or
116\,$M_{\sun}$\,Gyr$^{-1}$\,pc$^{-2}$) are found for Ring~1.  We
compare these values with Caldwell et al (1991) who found rates of $<$
0.4\,$M_{\sun}$\,Gyr$^{-1}$\,pc$^{-2}$ for early type spirals and
values of 4--20\,$M_{\sun}$\,Gyr$^{-1}$\,pc$^{-2}$ for Sc type
galaxies. The local SFR in the inner ring appears to be extremely
large, while those in Ring~2 appear to be similar to local SFR of Sc
galaxies.

\subsection{Azimuthal Variations in Ring~2}

Much of the foregoing argument for a threshold dependence of star
formation rate on ring strength is based on the assumption that all
three rings are generated by a similar dynamical process.  However, if
Ring~1 was formed, for example,  at an inner Lindblad resonance by some other
process, then our argument for a threshold star formation law would be
weakened. Nevertheless, there is evidence in Ring~2 alone, for
threshold behavior. We have already discussed in \S~7.2
the  unusually bright star forming complexes in the north-western
quadrant of Ring~2. We will now show that the
azimuthal variations in the strength of the star formation is very
similar to that expected from the models of Figure~4. In Figure~8 we
show an enlarged version of the Ring~2 data presented in Figure~7.
Unlike the earlier figure, Figure~8 connects the points with
increasing azimuth.  The number associated with each point represents
the knot number identified in Figure~2. It is clear from Figure~8,
that a strong azimuthal trend is evident as the star formation rate
increases in the north-western quadrant and these data resemble
the results predicted in Figure~4c. This systematic behavior in the
star formation rate as a function of azimuth provides a good
indication that the star formation rate is being controlled by a
global phenomenon and is not a stochastic process. This confirms our
earlier statement that the deviations of the \ion{H}{2} regions 19--23
from the linear correlation shown in Figure~5 are inconsistent with a
random noise process.

\subsection{The Size of the \protect{\ion{H}{2}} Regions}

There is an independent piece of evidence that a threshold process may
have changed the nature of the star formation in the overdense region
of the ring. In a study of \ion{H}{2} regions in galaxies of various
Hubble types, Kennicutt (1988) and Strobel, Hodge and Kennicutt (1991)
demonstrated that there was a well defined relationship between the
size of an extragalactic \ion{H}{2} region and its luminosity in the
H$\alpha$ line. Kennicutt chose to define the ``characteristic size''
of an \ion{H}{2} region in an external galaxy as the diameter of the
H$\alpha$ emission contour defined by the level 2$\times10^{-16}$
ergs\,s$^{-1}$\,cm$^{-2}$\,arcsec$^{-2}$.  It was shown that the size
increased monotonically with the flux from the \ion{H}{2} regions in
approximate accord with simple arguments concerning the size of the
Str\"{o}mgren sphere. In Figure~9, we show a contour map of the
H$\alpha$ emission at the level of 2$\times10^{-16}$
ergs\,s$^{-1}$\,cm$^{-2}$\,arcsec$^{-2}$.  What is striking about the
map is that the majority of the \ion{H}{2} regions which lie below the
``threshold'' value of R-band flux (Knots 6--16 in Figure~2) have
angular sizes which are barely or just resolved on at the level of a
few arcsec. However, as one approaches the north-western segment of
the Ring~2, the \ion{H}{2} regions suddenly grow in size until they
merge together to form one continuous structure which includes most of
the very active parts of the second ring and all of the first ring. It
appears that on the basis of the size of the \ion{H}{2} regions alone,
there is evidence for some form of transition from small to large
\ion{H}{2} regions which is consistent with our photometric result. Both
observational facts indicate that a threshold star
formation process is operating in this galaxy.

\subsection{Consequences for Other Star Formation Threshold Models}

We have presented evidence that suggests that the amplitude of the
stellar density wave in Ring~2 is a controlling influence in the
processes of star formation in the ring. At some critical (threshold)
surface density in the red starlight, a change occurs in the
properties of the \ion{H}{2} regions. The changes are consistent with
a sudden increase in star formation rate. We will now discuss possible
ways in which such changes may come about in the context of models of
star formation in external galaxies.

Kennicutt (1989) has presented convincing evidence for a star
formation threshold in a sample of normal disk galaxies. By combining
information about the total gas surface density derived from CO and
\ion{H}{1} measurements, Kennicutt was able to demonstrate that
massive star formation ($M > 10\,M_{\sun}$) occurs mainly in regions
of the disk which lie above a critical gas surface density. The form
for the critical density follows from a stability analysis of a thin
self-gravitating disk applied to the gaseous component (Toomre 1964,
Goldreich and Lynden-Bell 1965).  Kennicutt found that in regions
above the gas density threshold, the star formation law followed a
Schmidt type form ($\mbox{SFR}\propto\rho^{n}$ with $n\simeq1.3$),
whereas below the threshold, little or no massive stars were being
formed.

Firstly, we point out that our observations of Arp~10 cannot be
directly compared with the work of Kennicutt unless we make some
assumptions about the nature of the rings. We do not yet have
information about the distribution of \ion{H}{1} or CO observations in
this galaxy and therefore we cannot directly compare the star
formation rates in the rings to the gas surface densities. However, we
can appeal to our theoretical understanding of the ring galaxy
phenomenon to help restrict the range of possibilities.

If Arp~10 is a colliding ring galaxy, then the stellar surface density in
the ring will be a lower limit to the surface density in the gas
component. For example, in the absence of self-gravity in the gas, the
density profile of the expanding ring perturbation would closely track
that of the stars since both gas and stars are crowded into the
density wave (see SMA). In this simplified case, a measurement of the
old stellar surface density would be equivalent to a measurement of
the gas surface density in the ring. Even in the case where
hydrodynamic effects begin to dominate in the gas phase, the strength
of the stellar density wave will provide a lower limit to the density
in the gas. In general, whereas the gas will tend to occupy a smaller
volume than the old disk stars in the wave, the inverse is unlikely to
be true.

Since we do observe \ion{H}{2} regions in all three rings we conclude
that the rings must have already exceeded Kennicutt's critical density
over most of their area.  However, the marked increase of the H$\alpha$
flux in the north-west quadrant of Ring~2 suggests that in colliding
systems the star forming behavior is more complicated. In particular,
the existence of a threshold that depends on the strength of the
density wave may provide evidence for a breakdown of the linear
instability model which appears to provide a good description of star
formation in isolated galaxies. This may not be surprising, since
strong gravitational interactions and collisions are expected to drive
highly non-linear behavior in galaxies.

\subsection{Global Star Formation Rate in Arp~10}

Under the same assumption as in Section \S~7.1, we can calculate the
total SFR in the galaxy, including the two rings, the ring-arc and the
nuclear emission. The resulting SFR is approximately 0.94
stars\,yr$^{-1}$ (or 5.4 $M_{\sun}$\,yr$^{-1}$). Compared with a
typical late-type galaxy, this overall SFR is at the low end of the
distribution (GHT; Caldwell et al 1991).  This contrasts with the high
local SFR within the rings themselves. The apparent paradox is a
result of the narrowness of the rings. In comparison, a typical Sc
galaxy would have \ion{H}{2} regions of similar luminosity to Arp~10,
but would be distributed over a larger area of the disk.

\subsection{IMF Variations ?}

There is another way we can interpret the elevated H$\alpha$ fluxes in
the overdense regions without invoking a change in the star formation
rate. An equally plausible way of increasing the number of Lyman
continuum photons in the ``active'' regions would be to modify the IMF
in those regions. Since the number of Lyman continuum photons $N_{c}$
in all \ion{H}{2} regions is mainly produced by stars in the 30 to
60\,$M_{\sun}$ region, small changes in the slope of the IMF can lead
to quite large differences in the $N_{c}$. For example in order to
increase $N_{c}$ (and therefore by implication $\sigma_{H\alpha}$) by
a factor of 4--5 one only has to flatten the slope of the IMF from a
value of 2.35 to 1.95 (assuming an upper mass cut-off of
100\,$M_{\sun}$). This is sufficient to explain the sudden increase in
the H$\alpha$ emission in the overdense regions of Ring~2. However, if
the same process is to give rise to much brighter fluxes seen in the
inner ring, the slope would have to be modified substantially from
2.35 to 1.55. As discussed by Kennicutt (1983), such large changes in
the IMF slope would have noticeable consequences for the broad-band
colors of star formation regions. Since we do not yet have multicolor
images of Arp~10, the possibility that changes in IMF slope can
explain the higher H$\alpha$ fluxes in the overdense regions cannot be
ruled out. The slope of the IMF though, is not the only parameter that
can produce this result. If the upper mass cut-off is smaller than
60\,$M_{\sun}$, then this has a direct impact to the number of Lyman
continuum photons produced. It has been assumed that $M_{upper} =
100\,M_{\sun}$. However recent observations (Doyon, Puxley and Joseph
1992) indicate that there are galaxies with $M_{upper}<40\,M_{\sun}$.
It is not clear what physical mechanism could modify the IMF in the
dense parts of the ring-wave.

\section{Conclusions}

We have performed CCD imaging of the peculiar galaxy Arp~10 and have
drawn the following conclusions from our observations.

\begin{itemize}

\item Arp~10 is shown to contain two massive-star forming rings and an
outer star forming ring-arc indicating that star formation is
occurring concurrently on a variety of scales. The morphology of Arp
10 is consistent with that of a colliding ring galaxy.  We identify a
possible small elliptical companion galaxy which lies on the minor
axis of Ring~2, which may be the intruder galaxy.

\item The strength of the H$\alpha$ emission in the rings covers a
wide range of H$\alpha$ surface brightness, ranging from extremely
luminous in the inner ring (comparable to the most luminous \ion{H}{2}
regions in the late-type spiral galaxies) to rather faint \ion{H}{2}
regions in the outer ring-arc (comparable to the most luminous
\ion{H}{2} regions seen in Sa type galaxies).

\item We present evidence for a sudden change in the properties of the
ring \ion{H}{2} regions as a function of the strength of the
underlying stellar density wave. These are:

a) Taking the three ring structures together, there is evidence for a
discontinuous relationship between the logarithm of the H$\alpha$
surface brightness and the strength of the R-band continuum surface
density which support a threshold star formation model.  We interpret
the sudden increase in H$\alpha$ emission at a $\sigma_{R}$ of
21.3$\pm$0.2 mag\,arcsec$^{-2}$ as either a sudden increase in the SFR
or a change in the IMF at a critical matter density in the ring. In
either case, the result suggests that the star formation properties
are being controlled by the amplitude of the stellar surface density
in the ring.

b) The proposed critical threshold value in $\sigma_{R}$ falls within
the range spanned by Ring~2. At this apparent threshold level, the
star formation properties show a large dispersion, due mainly to an
increase by a factor of 5 in the star formation rates in the
north-western quadrant of the ring. This is the same quadrant of the
ring which was noted as unusual by Vorontsov-Velyaminov (1977). The
azimuthal variations of both star formation rate and R-band light show
remarkable similarity with model predictions for off-center ring
galaxy in which a threshold star formation law holds.

c) The size of the \ion{H}{2} regions are observed to grow suddenly
in the region of the enhanced star formation discussed in b) above,
again supporting the proposed threshold picture.

\item The results indicate that the star formation behavior may
differ from that of a normal quescient disk system as a result of the
existence of strong non-linear waves induced in the disk by a
collision. A full comparison with the gas density threshold models of
Kennicutt (1989; 1990) will require observations of the cool gas content of
the three star forming rings.
\end{itemize}

\acknowledgments

We thank C. Struck-Marcell (Iowa State University) for stimulating
discussions and L. Fontaine (Graphic Design, ISU) for advice and help
regarding Figure~4. The authors would like to thank J. J. Eitter (Iowa
State University) for invaluable help at all stages of the
observations at the Fick Observatory, and G. Aldring (U. of Minnesota)
for providing the design for the f/4 focal reducer. PNA wishes to
thank the Department of Physics and Astronomy for providing travel
funds to KPNO. The authors are grateful to an anonymous referee for
suggesting some significant improvements to the paper.


\appendix
\begin{center}
{\bf Appendix}
\end{center}
In this section we compare the R-band image of Arp~10 taken at KPNO
with the one that we took at Fick Observatory, as a further check that
it has no contamination by hydrogen emission lines. Since the two
images had different resolution we convolved them with a gaussian
profile down to an effective resolution of FWHM\,$= 3.3$\,arcsec. A
contour graph of each image is presented in Figures~10 and 11. Our
calculations of the intensity distribution along Ring~2 in both images
showed that there was no discernible difference between the two, down
to a level of $<$\,5\%

%
%

\clearpage

\clearpage

%
%

\clearpage

%
%

\begin{figure}
\caption{Greyscale image of Arp~10 through the R-band filter. Note the
possible elliptical companion galaxy seen to the north-east of Arp~10
and the bright extra-nuclear knot which lies to the west of the
nucleus. North is to the left and west is at the bottom in all
the images in this paper.}
\end{figure}

\begin{figure}
\caption{Greyscale image of Arp~10 through a narrow band (80 \AA\
wide) filter centered close to the wavelength of redshifted H$\alpha$
and [\protect{\ion{N}{2}}].  The inset shows the same region defined
by the box but at a different greyscale contrast level, emphasizing
the ring-like nature of the inner ring. The numbered region are
discussed in the text. }

\end{figure}

\begin{figure}
\caption{Contour map of Figure~2. The contour levels are at
0.36$\times$10$^{-16}$ ergs\,cm$^{-2}$\,arcsec$^{-2}$ increments, between
0.36$\times$10$^{-16}$ ergs\,cm$^{-2}$\,arcsec$^{-2}$ and
14.3$\times$10$^{-16}$ ergs\,cm$^{-2}$\,arcsec$^{-2}$. The inset shows
details of the inner ring and nucleus.}
\end{figure}

\begin{figure}
\caption{(a)The gas density distribution for an off-center ring galaxy
model (Model E) from the work of Appleton \& Struck-Marcell (1987).
Note that the density varies with azimuth around the ring. We have
labeled azimuthal positions around the ring with the letters A to F
for reference. (b) The relationship between the star formation rate
(SFR) and gas density for the model shown in (a). The behavior follows
that of a Schmidt law (see text). The letters refer to the azimuthal
positions given in (a).  (c) The same relationship as in (b)
but for a ``threshold'' star formation law. Here the SFR follows a
Schmidt law below a critical density (In this case 1.5 x the
unperturbed density) and then changes to a steeper form in which SFR
$\propto \rho^{6}$ above the critical density. }
\end{figure}

\begin{figure}
\caption{Plot of H$\alpha$ versus R-band surface brightness for the constant
isophote method (see \S~4, Table~1). The different symbols indicate
data for the three ring structures. Note that the large error bars for Ring
1 indicate our uncertainty in the removal of the bulge light which
dominates the R-band emission for this inner ring. The solid line
shows a linear fit to these data and the dotted lines
show the result of a two line fit (see text).}
\end{figure}

\begin{figure}
\caption{Plot of the residuals from Figure~5 after the subtraction of
the linear fit through the points of Ring~2 and Ring-arc~3.}
\end{figure}

\begin{figure}
\caption{Plot of the SFR in the three ring structures, in units of
stars\,Gyr$^{-1}$\,pc$^{-2}$, versus the R-band surface brightness for
the constant isophote method (See \S~4, Table~1). Note that the
uncertainties in the Ring~1 points are identical to those shown in
Figure~6.}
\end{figure}

\begin{figure}
\caption{The azimuthal behavior of the SFR and R-band continuum for
Ring~2. The numbers refer to the H$\alpha$ emitting regions identified
in Figure~2 and increase numerically as a function of increasing
position angle around the ring. We observe the striking similarity with the
threshold model shown in Figure~4c.}
\end{figure}

\begin{figure}
\caption{A contour map of the H$\alpha$ emission at the level of
2$\times$10$^{-16}$ ergs\,cm$^{-2}$\,arcsec$^{-2}$. This isophotal
level was used by Kennicutt (1988) to define the ``characteristic
size'' of an \protect{\ion{H}{2}} region. Note that the
\protect{\ion{H}{2}} region complexes suddenly grow in size in the
bright (north-west) region of Ring~2.}
\end{figure}

\begin{figure}
\caption{A contour map of the KPNO image presented at Figure~1 after
convolution with a gaussian profile. The resulting effective
resolution is 3.3\,arcsec (FWHM).}
\end{figure}

\begin{figure}
\caption{Contour map of the Fick Observatory CCD image obtained
through a narrow-band filter. The filter is shifted with respect to
H$\alpha$ emission line and as a result is completely free of any
possible line contamination (see text).  Note the strong similarities
(down to a level of $<$\,5\%) between this image and Figure~10.}
\end{figure}

\end{document}